# Big Data Analytics in Cloud environment using Hadoop

Mansaf Alam and Kashish Ara Shakil

Department of Computer Science, Jamia Millia Islamia, New Delhi

**Abstract:** The Big Data management is a problem right now. The Big Data growth is very high. It is very difficult to manage due to various characteristics. This manuscript focuses on Big Data analytics in cloud environment using Hadoop. We have classified the Big Data according to its characteristics like Volume, Value, Variety and Velocity. We have made various nodes to process the data based on their volume, velocity, value and variety. In this work we have classify the input data and routed to various processing node. At the last after processing from each node, we can combine the output of all nodes to get the final result. We have used Hadoop to partition the data as well as process it.

## 1. Introduction

There is a rampant increase in the amount of data being produced from varied sources. This can be attributed to the instrumentalisation of the current society and personnel's leading to storage and production of vast amounts of data. Since, the data being produced is huge with a lot of variety and the rate of production is also rapid. Thus, the traditional systems fail to manage this data and this is what led to the buzz word called Big Data. Big Data is a term which refers to the explosion of variety of data produced from disparate sources [1]. It is characterized by five features or attributes i.e. high volume, variety, veracity, visibility and velocity. Since, this kind of data is beyond the management scope of traditional systems therefore in order to mine such kind of data we need analytics' solutions that can help in gaining insights from both structured and unstructured data.

At present scenario its instrumental to blend both big data and analytics into a single entity termed as big data Analytics. Analytics involves examination of data to derive meaningful insights such as hidden patterns and trends that can in turn benefit the organizations in making important business decisions and developing newer business models. The problem of data deluge imposes potential challenges involved in processing and extracting useful information from data. It also requires skills for management and analysis of huge data sets.

Cloud computing serves as a quintessential solution for handling big data and hosting big data workloads. Cloud computing has revolutionized the way in which computing resources can be utilized by providing facilities such as pay per use, rapid elasticity and dynamic scalability. It provides the users with an illusion of infinite storage and compute capacity. The cloud resources can be used in private mode through private cloud or can be shared publicly using a public cloud such as Amazon EC2 and Microsoft Azure. Cloud therefore serves as a scalable technology with low upfront investment costs. Thus, the proposition value associated with using cloud as a platform for carrying out analytics is quite strong and therefore it is well suited for carrying out scalable data analytics.

Hadoop is a technology that can be used for handling big data. It can play a significant role in opening gates to new insights out of data and can easily handle flood of huge unstructured data sets coming from sources such as sensors, mobile devices and social media.

This paper presents about how hadoop can be used as technology on cloud for meeting the big data needs of users and discusses about the proposed hadoop based workflow for handling big data. We also present a case study of analysis carried out on movie data for mining many useful information from it which includes finding the number of movies released between a given period and the number of movies having a certain rating besides other information's.

The rest of this paper is organized as follows: Section 2 presents a survey of the related approaches used for big data analytics, Section 3 discusses about hadoop as a platform for meeting the big data needs and requirements. Section 4 shows our proposed workflow for carrying out big data analytics. Furthermore, Section 5 discusses our case study for analytics of movie data. Finally the paper concludes with conclusion and future directions in section 6.

## 2. Related Work

In the research paper [6], the researchers have discussed the assisting developers of BDA Apps for cloud deployments. In their paper they have proposed a lightweight approach for uncovering differences between pseudo and large-scale cloud deployments, their approach makes use of the readily-available yet rarely used execution logs from these platforms. They have done a case study on three representative Hadoop-based BDA Apps and have shown that their approach can rapidly direct the attention of BDA App developers to the major differences between the two deployments. The research work [7] outline challenges in analyzing big data for both *data at rest* and *data in motion*. They have described two kinds system for big data which is at rest namely NoSQL systems for interactive data serving environments and systems for large scale analytics based on MapReduce paradigm, such as Hadoop, The have discussed that the NoSQL systems are designed to have a simpler key-value based data model having in-built *sharding*, so this work in a distributed cloud based environment. In contrast, to run long running decision support and analytical queries consuming and possible producing bulk data by use Hadoop based systems. For processing data in motion, they have present use-cases and illustrative algorithms of data stream management system. In a research paper [8] explained that an thought-provoking thing of the cloud paradigm the cost of using 1000 machines for 1 hour, is the same as using 1 machine for 1000 hours, these implies that a Hadoop job's performance can potentially be improved, while incurring the same cost, this proved Hadoop is put organized to feat parallelism.

## 3. Hadoop

MapReduce [2] paradigm provides a highly scalable and flexible platform for managing high scale data sets. The crux of MapReduce lies in its ability to distribute data into different nodes and process it in a parallel fashion which is not transparent to its users. Hadoop [3] is an open source implementation of MapReduce [5] which has gained immense popularity amongst different organizations such as facebook, Yahoo and twitter for running their data intensive applications. Its popularity can be attributed to its ability to provide a highly scalable, parallel and fault tolerant framework that supports automatic distribution and computation across a cluster of commodity hardware [4]. Thus, hadoop is well suited as a platform for carrying out large scale data analysis.

Hadoop supports the MapReduce programming framework. Figure 1 presents an overview of Hadoop MapReduce framework. In this framework programming is divided into two phases Mapper Phase and Reducer Phase. Firstly the data is divided into fixed sized blocks depending upon the block size settings. Each of these blocks of data is processed at different nodes . At each of these nodes Mapper function is run for carrying out processing. The data is processed at each of the Mapper nodes . The results produced at each of the Mapper nodes are than combined using reduction phase. The combined results are than generated as the output.

Some of the key characteristics that led to the rapid popularity of hadoop are:

a. Economic benefits: Since hadoop is an open source framework which works on cluster of commodity hardware. Therefore, a hadoop cluster can be set easily without any initial capital investments.
b. Scalability: Hadoop provides support for dynamic scalability, any amount of data can be stored and processed using hadoop framework by addition or removal of machines depending upon the user requirements.

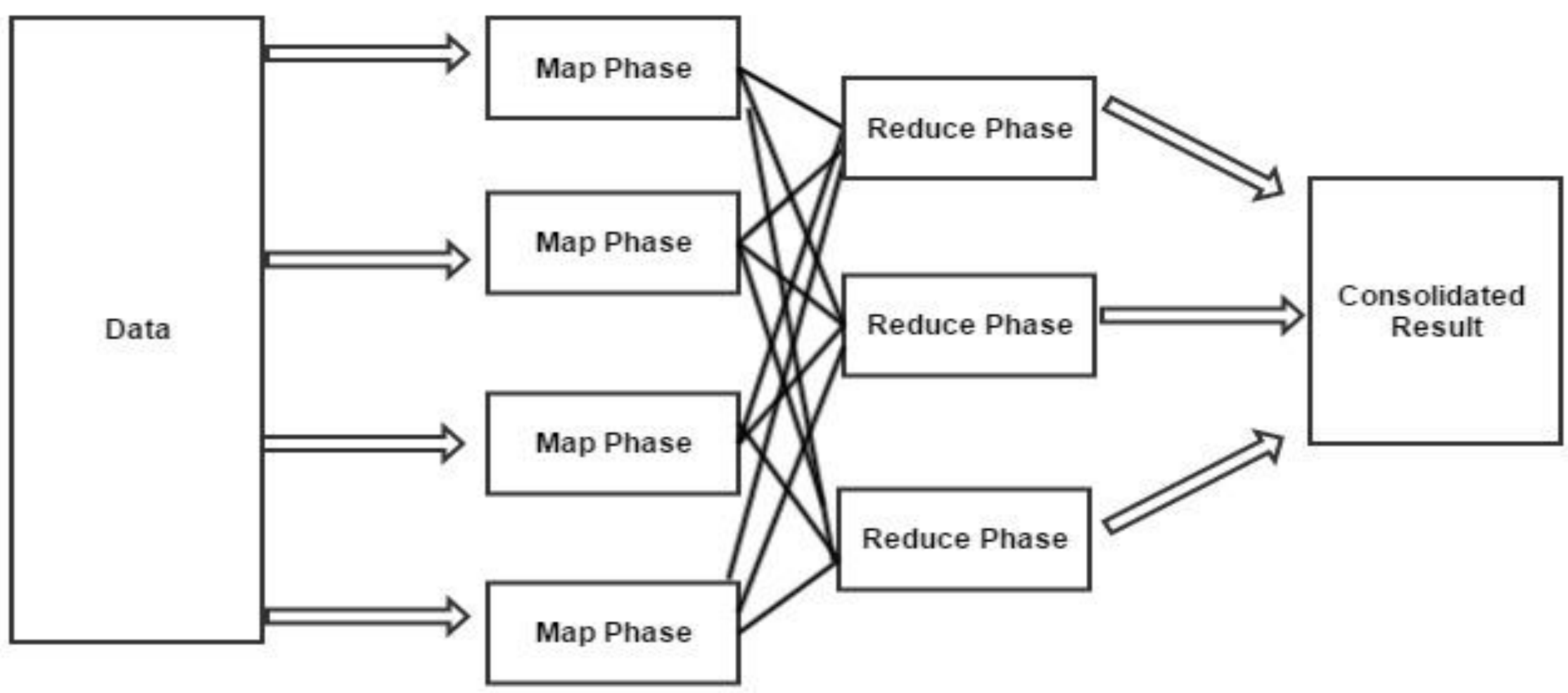

**Figure 1: Hadoop MapReduce model**

c. Reliability: In hadoop by default three copies of a dataset is made i.e. by default replication factor of a hadoop job is three. This feature offers its users with a reliable and robust framework because even if one of the machines holding that data set goes down the system will still be running as the data will be available at other replicated locations.
d. Flexibility: hadoop offers its users with a highly flexible framework. The number of machines can easily be added and removed from the cluster depending upon the user requirements.

4. Hadoop in Cloud Environment

## 5. Big Data Analytics in Cloud Environment

### 5.1 Proposed Workflow

The data are coming from various sources with different nature of data as well as in large volume of data. All data coming from different source are big data because volumes of data are very high, flow of data is at very large speed. There is variety in data and the data is coming is very important. In our proposed work, we have first route the all data coming from different sources to cloud environment and store it in HDFS of Hadoop framework. The Big data stored in HDFS is the transfer to classifier module which is responsible to classify the coming data and send it various designated node for processing the data. We have also proposed a algorithm to classify the coming data from different source and route it to various des8ignated node processing. After processing from different designated node, the result produced by designated nodes are transfer to a result module and all result from designated node are combined together and final result is then send to client as well as the result is also sent to HDFS of Hadoop Framework to stored it for further use. The proposed concept of bog data analytics in cloud environment using Hadoop framework is shown in Figure 2. Let us consider s is different sources of data and n is number of total sources. When data is transfer to cloud and Hadoop, due to heavy transmission of data from different sources, it may be propagation delay, assume propagation delay is $\partial$p and the transmission time of data from source to processing unit is p. Due to propagation delay some processor node to wait for some unit of data, waiting time for processing unit for remaining data for a job is $t_w$ . There are m processing unit in the entire system, which is capable to process m jobs at time in parallel, the processing unit take time $t_p$ to process a single job. The total time to process a sing job is T. The T is calculated using equation 1.

$$\boldsymbol{T = \partial p + p + t_p} \quad \textbf{(1)}$$

The equation calculates time for processing single job if there is delay in data arrival.

If there is no delay in arrival of data at processing unit then the time required to process a sing job is calculated the with equation 2.

$$\boldsymbol{T = p + t_p} \qquad \mathbf{2}$$

There is a situation in which there is some times propagation delay and some time there is no propagation delays in data from source to processing unit, so in general calculate average processing time. The average processing time is calculated with the help of equation 3. Let us consider average time to process a single job is $A_{t.}$

$$\boldsymbol{A_t = \frac{(p + t_p) + (\partial p + p + t_p)}{T}}$$

$$\boldsymbol{A_t = \frac{2p + 2t_p + \partial p}{T}}$$

$$\boldsymbol{A_t = \frac{2(p+t_p)+\partial p}{T}} \qquad \mathbf{3}$$

**Definition 1:** The average time required to process single Job in Hadoop based cloud environment is $A_{t,}$ where $A_t$ is defined as the sum of time with delay and time without delay per unit of total time.

The above equation calculates average time for single processing unit. now we consider entire system**.** The propagation delay of entire system including all processing unit is $P_e$ .

$$\boldsymbol{P_e = \sum_{i=0}^{m} \partial p_i} \qquad \mathbf{4}$$

The time required to process the entire job by all processing unit of the system with propagation delay is $T_{\partial e}$.

$$\boldsymbol{T_{\partial e} = P_e + \sum_{i=1}^{m} p_i + \sum_{i=1}^{m} t_{pi}}$$

Let consider there is no delay in the entire system while processing the incoming data, then the time required to process the incoming data is $T_{e.}$

$$\boldsymbol{Te = \sum_{i=1}^{m} p_i + \sum_{i=1}^{n} t_{pi}}$$

The average time required to process the entire job entire processing unit in the proposed system is $A_e$

$$A_e = \frac{P_e + \sum_{i=1}^{n} p_i + \sum_{i=1}^{n} t_{pi} + \sum_{i=1}^{n} p_i + \sum_{i=1}^{n} t_{pi}}{\sum_{i=1}^{n} \quad T_i}$$

$$A_e = \frac{P_e + 2\sum_{i=1}^{n} p_i + 2\sum_{i=1}^{n} t_{pi}}{\sum_{i=1}^{n} \quad T_i}$$

$$A_e = \frac{P_e + 2(\sum_{i=1}^{n} p_i + \sum_{i=1}^{n} t_{pi})}{\sum_{i=1}^{n} \quad T_i} \qquad 5$$

**Definition 2: The average time required to process entire jobs in entire system in Hadoop based cloud environment is** $A_e$

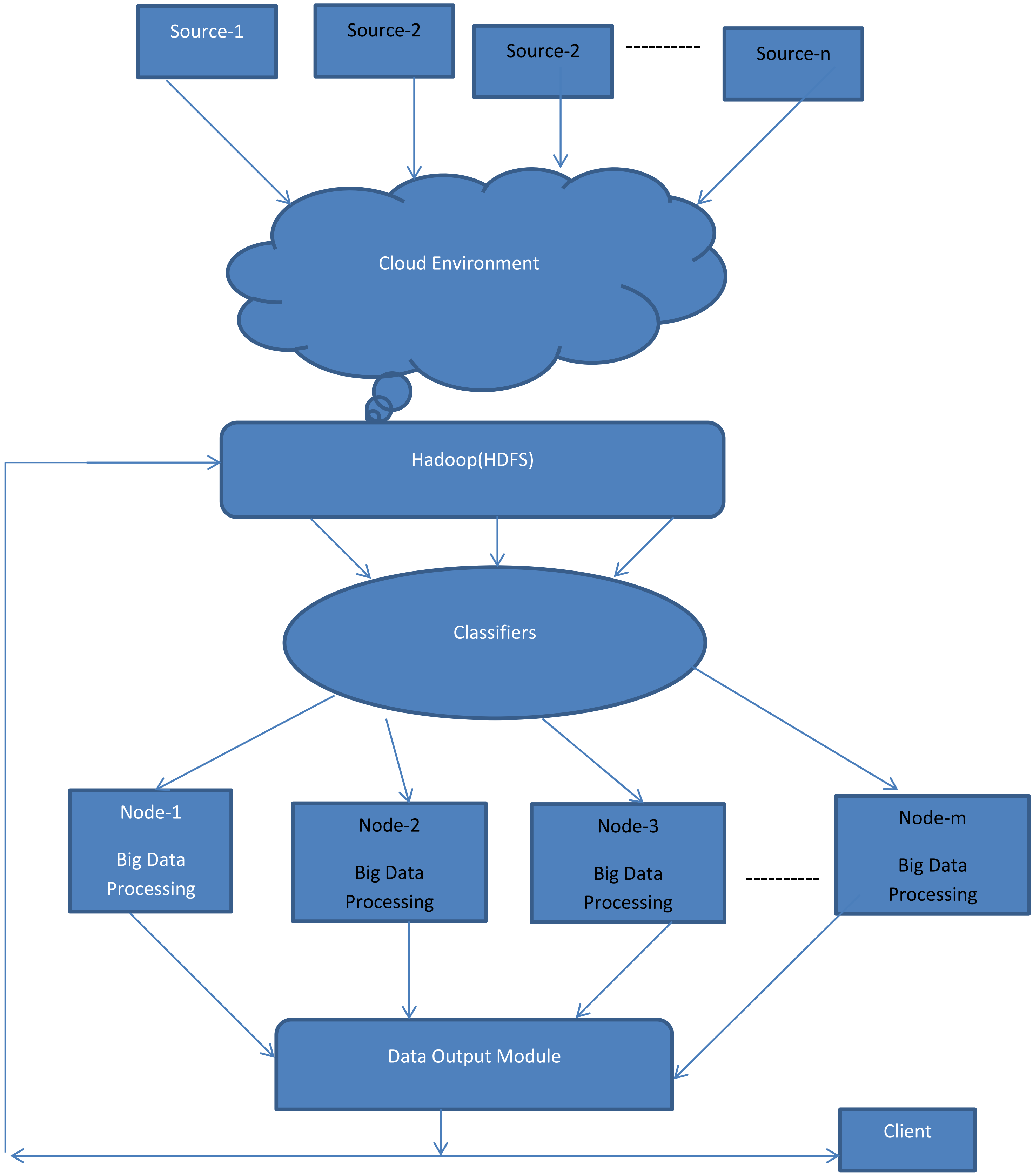

**Figure 2:** Proposed system to for Big Analytics in Cloud Environment using Hadoop

## 6. Case Study

In order to further elucidate the proposed workflow we took up a case study of analytics for video data. Video data management usually involve processing of large amount of data to gather information such as year of release of a particular movie and the ratings awarded to it. Since there is a large number of movies releasing each year thus its a very challenging task to effectively store and efficiently process this data. This challenge can be easily met by making use of hadoop based cloud workflow proposed in section 4.apig has been used by us as a tool for meeting our analytics endeavors. The various modules in our case study include the following:

A. Distributed Storage

The data was processed using hadoop it was initially imported into hadoop distributed file system(HDFS). HDFS uses master slave architecture which includes a single name node and several data nodes. The name node acts as a central sever and manages the entire file system and access to blocks of data lying on the HDFS. The actual data is stored at the data node which acts the slave node. The advantage of using HDFS as a storage system is that it provides dynamic scalability of the system i.e. it can scale up to several nodes depending upon the user requirements. Figuren1. shows how data has been imported in the proposed workflow into the grunt shell for carrying out analysis.

B. Data Processing Module:

This module consists of individual data nodes. Each of these data node is responsible for carrying out execution in terms of Mapper functions. The advantage of carrying out processing using MapReduce paradigm is that there is very little overhead involved in movement of data due to features such as rack awareness offered by the MapReduce paradigm.

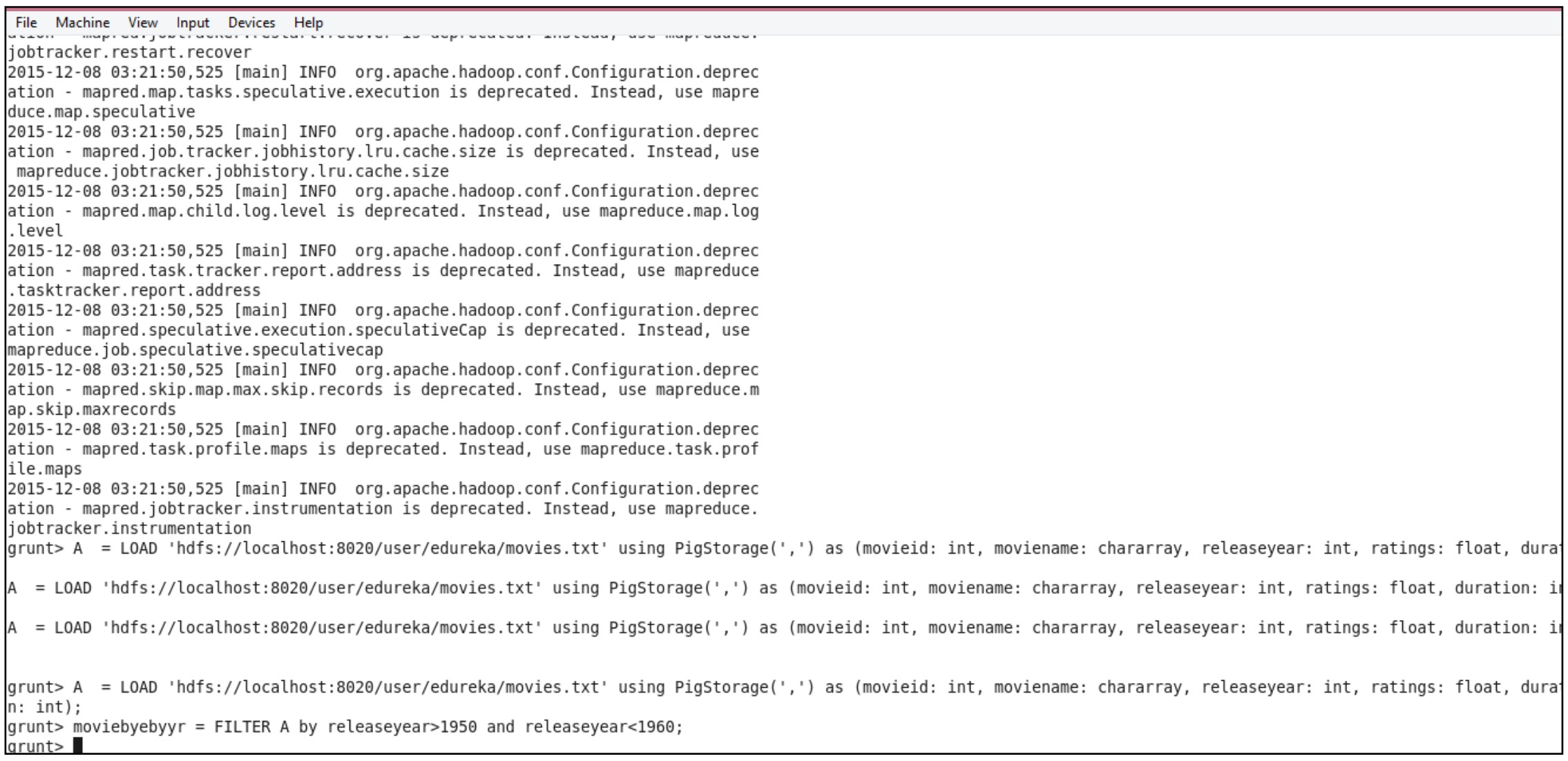

Figure 2: Data Import from HDFS to pigs grunt shell

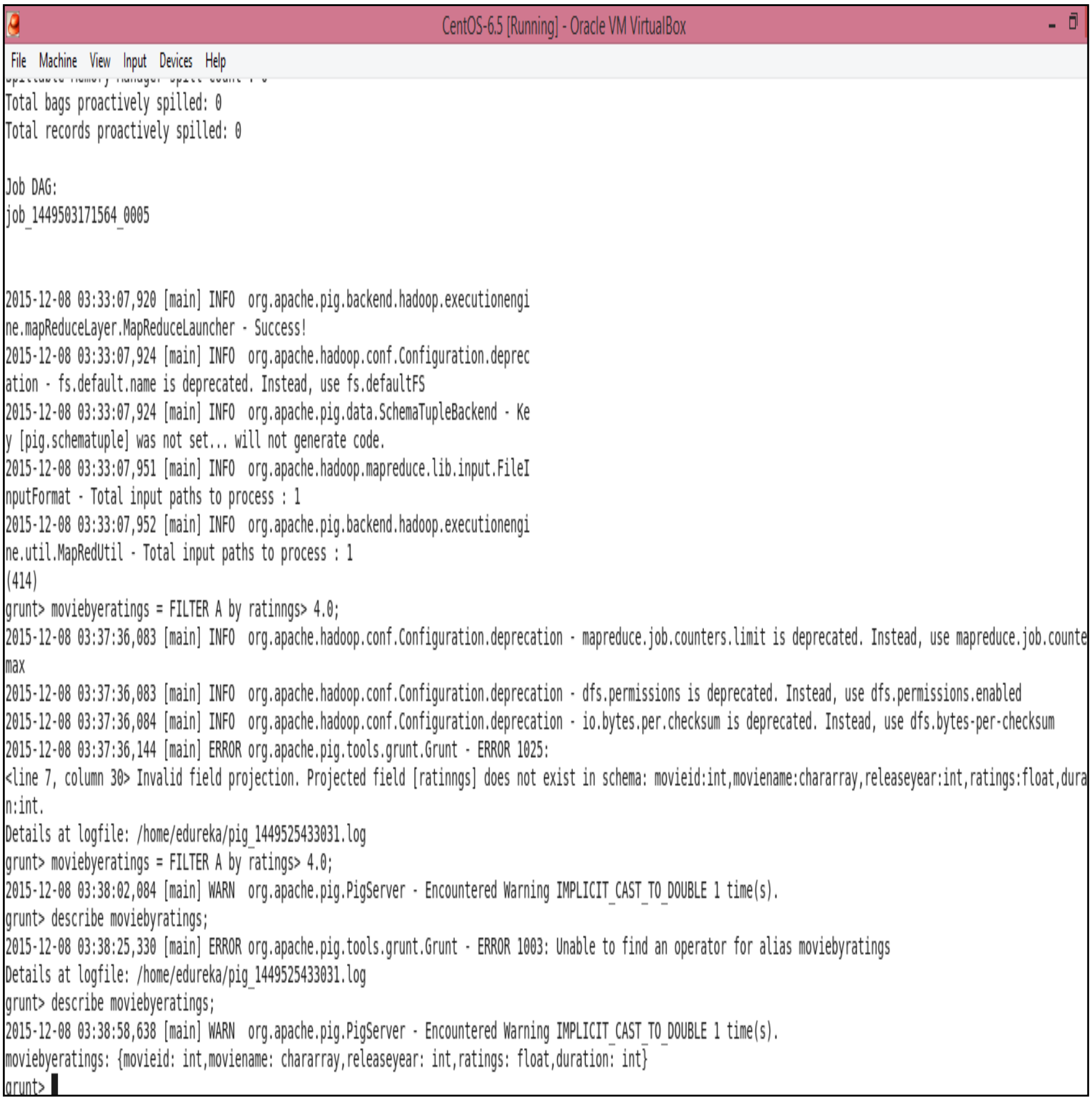

Figure 3: Sample Example of pig script

In order to carry out processing in our case study we performed operations like finding the number of movies released in a decade. For example figure 3 shows the pig script used by us for finding the number of movies having rating more than 4.

C. Data Output Module:

In order to receive output of our movie data analytics the results are obtained by the reducer by combining together data processed at different Mapper nodes. The reducer then provides a consolidated output to the user. This entire process of breakage of data into different blocks and processing at each node is carried out in a manner which is not transparent to its users. Figure 4 shows the result of output produced by the query fired in figure 3.

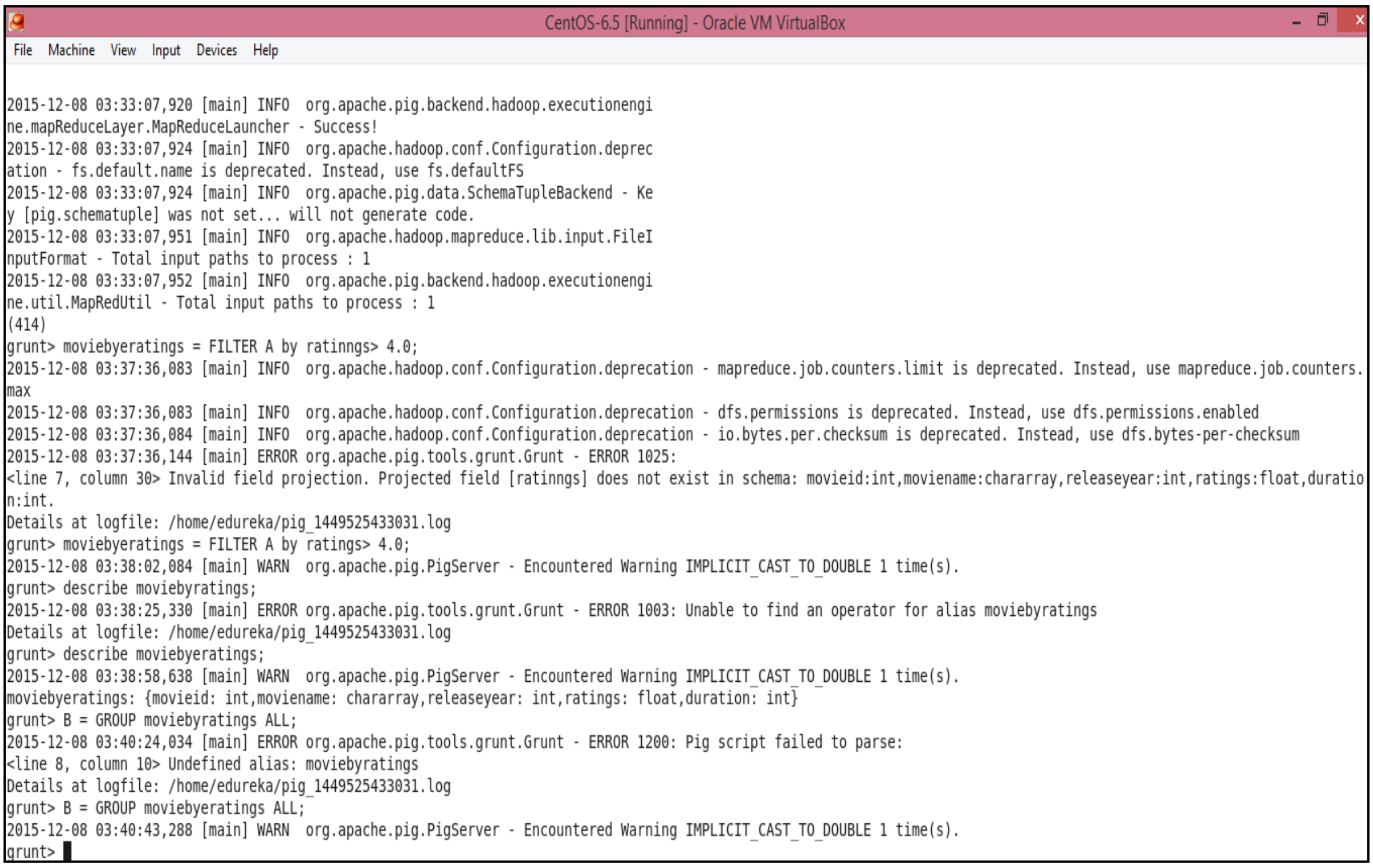

Figure 4: Output of pig script fired in figure 3

## 7. Conclusion and future directions

We can manage huge amount of data as well as big data can be managing in very easy ways in less amount of time. In our research work we found that the average processing time very les while processing in the cloud environment. We have done a case study, we have taken movie data and analyze it and investigate time required to process the data. In future a classifier is will be more flexible and scalable to get more accurate result.

.